\documentclass[12pt,a4paper]{article}
\usepackage{graphicx}

\begin{document}

%
%

\title{Optimal signals for spectral measurements at EKB SuperDARN radar: Theory and Experiment}

%
%

%
%

\author{O.I. Berngardt,
A.L. Voronov,
K.V. Grkovich}









%
%

\maketitle

\begin{abstract}
The requirements for an optimal sequence for SuperDARN radar are stated.
Obtained quasioptimal signals up to 26-pulse sequence are based on optimal
and nearly optimal Golomb rulers. It is shown that standard 7-pulse
SuperDARN sequence is one of the quasioptimal sounding signals, but
8-pulse kascan SuperDARN signal is not. The characteristics are calculated
for quasioptimal sounding sequences up to 26-pulse one. 
It is shown that the most effective signals for the spectral measurements are
10, 12, 14, 18, 20, 24 and 26 pulse sequences. 

We present the results
of the first spectral measurements at the EKB SuperDARN radar with
quasioptimal 8- and 9-pulse sequences, as well as with the optimal 10-and 12-pulse
sequences. A comparison with the results obtained with standard 7-pulse SuperDARN
sequence were made. The continuity of results in amplitude and velocity
is shown, the expected improvement in the spectral width are also demonstrated.

\end{abstract}


\section{Introduction}

The main objective of SuperDARN radars is to study the spectral characteristics
of ionospheric irregularities elongated with the Earth magnetic field, 
that allows one to study the dynamics of ionospheric electric fields
at high latitudes. It should be noted that the signal scattered by
elongated irregularities can be divided into 4 main types \cite{Haldoupis_1989},
wherein the separation is carried out on the details of their spectral
characteristics. However, the study of the dynamics of these irregularities
is practically impossible without a detailed analysis of the spectral structure
of the scattered signal \cite{Hanuise_et_al_1993, Schiffler_et_al_1997, Danskin_et_al_2004}.

The main difficulty of using SuperDARN radars for studying the spectral
properties of the scattering irregularities is the shape of the sounding
signal. Currently for SuperDARN the two types are used: standard 7-pulse \cite{Greenwald_et_al_1983}
 and 8-pulse katscan sequences \cite{Riberio_et_al_2013}.
The use of these sequences for the detailed study of the spectral
characteristics is almost impossible, since the spectral resolution
provided by these sequences is too poor. It is equivalent to the 100-200
m/s for the speed.

At the moment, the problem of low spectral resolution on SuperDARN
radars is solved by complex techniques that allow one to estimate
the Doppler drift and the average spectral width of the signal without
obtaining the spectrum, but based on the analysis of the correlation
function and its phase structure, as well as on the basis of model
assumptions \cite{Hanuise_et_al_1993, Schiffler_et_al_1997, Barthes_et_al_1998, Chisham_et_al_2007, Riberio_et_al_2013}.
The fine structure of the spectra is frequently finer than 200m/s \cite{Huber_and_Sofko_2000, Danskin_et_al_2004},
so for stable recognition and investigation of these structures the standard measurments require 
improving spectral resolution at least 2-3 times or use model-based inversion techniques \cite{Schiffler_et_al_1997, Barthes_et_al_1998}.

\begin{figure}
\includegraphics[scale=0.5]{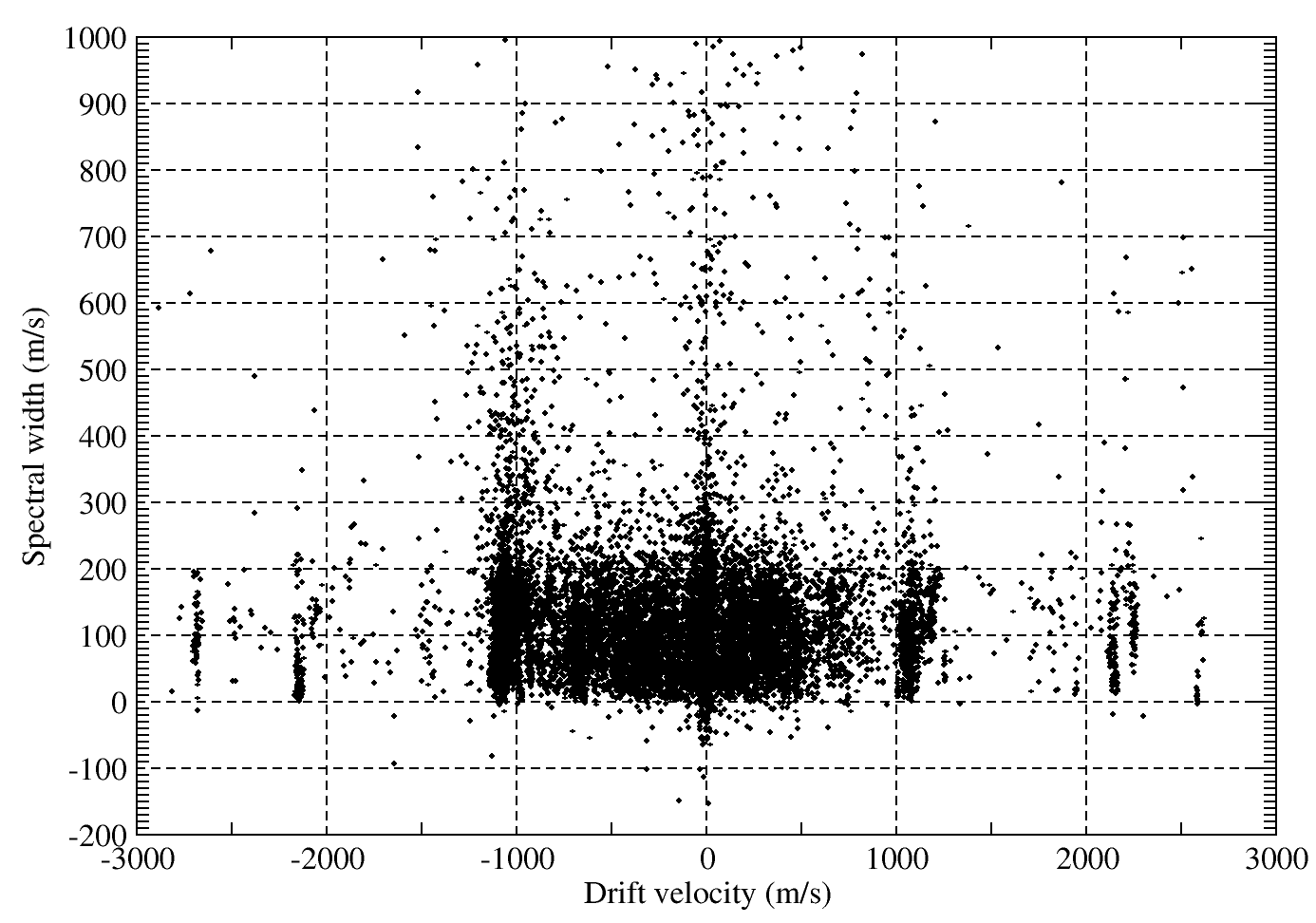}

\caption{One of the typical distributions of the measured parameters
for EKB SuperDARN radar (procedure for determining parameters
- FitACF)
}
\label{fig:1} 
\end{figure}

It is clear that in some cases the model representations of the scattered
signal are not in agreement with experiment, and in this case event
leads to significant errors in the estimation of parameters. These
errors may lead for example to the appearance of paradoxical negative spectral
widths. An example of the signal distribution over the velocities
and the spectral width at EKB radar is shown at Fig.\ref{fig:1}.

Impossibility of calculating the spectra of SuperDARN radar signals
leads to complications in recovery methods for spectral parameters
of the irregularities. At Fig.\ref{fig:1} one can see one consequence of
these errors - the negative spectral widths calculated by FitACF program.
These errors are associated with the processing algorithms for correlation
functions \cite{Riberio_et_al_2013} and caused by the inability to
restore the spectrum of the scattered signal from sounding results with 
standard sounding signals.

Thus, the problem of using optimal sounding signals in the SuperDARN
technique has the one of the important values \cite{Greenwald_et_al_1983, Schiffler_1996, Greenwald_et_al_2008}. 
Feature of SuperDARN
radars functioning is the requirement that the product of the spectral
resolution and spatial resolution to be significantly less than the
speed of light in vacuum. This means that the use of simple pulse
signals (for which this product is equal to the half of velocity of light)
in this problem is impossible and one needs the use of special signals
in conjunction with additional approximations depending on the nature of the
scattering.

Sequences and processing methods that implement these special signals,
have been well studied in the incoherent scattering technique: these
are multipulse sequences \cite{Farley_1972}, random phase codes \cite{Sulzer_1986},
alternating codes \cite{Lehtinen_and_Haggstrom_1987}
(including the polyphase ones \cite{Markkanen_et_al_2008}) and the
technique of effective subtraction \cite{Berngardt_and_Kushnarev_2013}.
However, in the SuperDARN technique the product of necessary spectral
resolution and range resolution is so small that one can effectively
use only multipulse sequences. In this case, the signal reception
is carried out by means of the transmission of the short pulses, which
form the sounding sequence. Therefore, SuperDARN
radars currently used only multipulse sounding sequences that have
necessary characteristics - 7-pulse standard sequence \cite{Greenwald_et_al_1983}
and 8-pulse katscan sequence \cite{Riberio_et_al_2013}, implementing
the principle described in \cite{Farley_1972}, as well as 13-pulse
sequence based on mixing forward and reverse optimal Golomb sequences\cite{Greenwald_et_al_2008}.

\section{Quasioptimal and optimal SuperDARN sounding sequences}

The convenient basis for the understanding of backscattering techniques
is the concept of the weight volume \cite{Potekhin_and_Berngardt_2000,Berngardt_and_Potekhin_2002,Berngardt_and_Kushnarev_2013}.
Weight volume shows how much of the correlation function of the received
signal is related to the correlation function of inhomogeneities,
and corresponding set of ranges. The relationship between the scattering cross
section of irregularities $\sigma(r,T)$ and the average correlation
function of the received signal $P(T,T_{0})$ measured in the moment
$T_{0}$ after transmitting sounding signal in a first approximation
is given by the expression \cite{Berngardt_and_Kushnarev_2013}:

\[
P(T,T_{0}) \sim \int W(r-cT_{0}/2,T)\sigma(r,T)\frac{dr}{r^{2}}
\]
where c - speed of light. 

The shape of the weight volume $W(r,T)$ with standard correlation
processing technique is determined by the shape of the sounding signal
envelope $a(t)$:

\[
W(r,T)=a(rc/2)\overline{a(rc/2+T)}
\]

The spatial resolution varies for each of the delay $T$ of the correlation
function, and is determined as the width of the weight volume at the
fixed delay $T$. From the structure of the weight volume it is obvious
that the spatial resolution at zero delay ('lag') $T=0$ is determined
by the total duration of the signal and can not be improved (except
in special subtraction techniques like \cite{Berngardt_and_Kushnarev_2013}).
However, the spatial resolution at the other lags $T$ depends of
the signal shape. Resolution over the lags is inversely proportional
to the spectral resolution and is equal to the total duration of the
signal.

For SuperDARN sounding signal \cite{Greenwald_et_al_1983, Riberio_et_al_2013},
that is a sequence of $N$ elementary almost rectangular pulses $T_{p}$
(300us), spaced by multiples of $T_{ip}$ (n * 2.4ms) and having total
duration $T_{sec}$ (64.8ms), we can define a spatial resolution
of $d(T)$ as width of weight volume for a given delay:
\[
r_{avrg}(T)=\frac{\intop rW(r,T)dr}{\intop W(r,T)dr}
\]

\[
d(T)=\left(\frac{12}{c^{2}T_{p}^{2}}\frac{\intop\left(r-r_{avrg}(T)\right)^{2}W(r,T)dr}{\intop W(r,T)dr}\right)^{1/2}
\]

Property of the function $d(T)$ is that it reaches the value 1 at
the 'good' lags, and 0 - at 'bad' lags. 

If $d(T)>1$  then  this lag is with
poor spatial resolution larger than the duration of a short elementary
pulse $T_{p}$, we will not take such sounding signals into consideration. 

Therefore, the for synthesis of optimal signal, the requirements 
are easy to formulate as the requirements to $d(T)$  (and through
it - to the weight volume and to the sounding signal), which is formed
by the shape of the sounding signal.

In the case of the SuperDARN technique the requirements for optimal
signal are following: 

1. The weight volume is such that at all the lags, except zero one,
the spatial resolution for all the lags that are multiples of the
elementary interpulse period is 'excellent' - not less than that of a single pulse:

\[
T>T_{ip},\, d(T=nT_{ip})\leq1
\]

The requirement of the excellent spatial resolution for each point is based on the following qualitative consideration.
In case some of the points are made with worse spatial resolution, the energy of the weight volume for this lag is lost for analysis, 
so in this case this part of the energy of transmitted signal is used not for obtaining correlation function. As a result using signals,
having points with spatial resolution worser than single pulse will decrease actual signal-to-noise ratio and require additional accumulation. 
Examples of pulses having points with worse spatial resolution can be found, for example, in \cite{Schiffler_1996, Greenwald_et_al_2008}.

2. Lags $T=nT_{ip}$ at which the spatial resolution is equal to
the elementary pulse one ($d(T=nT_{ip})=1$), are considered to be
good ('good' lags), and at which the correlation function of the scattered
signal does not correspond to the scattering from the medium ($d(T=nT_{ip})=0$)
- are considered to be bad ('bad' lags).

3 The ratio $\xi=M/L$ of good lags number $M$ in the correlation
function

\[
M=\begin{array}{c}
L\\
\sum\\
n=1
\end{array}\delta_{d(nT_{ip}),1}
\]

to the total number of lags $L$ in the correlation function

\[
L=T_{sec}/T_{ip}
\]

must be maximized for a given number of pulses in the sequence.

Requirements and definitions 1-2 are known and used long time, requirement
3 (optimality criterion for sounding sequence) introduced for
reasons of most efficient use of the received data - the highest possible
number of lags in correlation functions of received signal should
be used for processing, the same considerations were used in \cite{Farley_1972}.

To find the optimal sounding signals, we used a parallel algorithm
for search of optimal sequences with maximal $\xi$ within the class
of sequences with a fixed number of pulses. The search was made by
using computing clusters ISC SB RAS 'Blackford'
and 'Academician V.M.Matrosov' (http://hpc.icc.ru/)
by Monte Carlo technique, modified for taking into account the peculiarities
of the desired sequences.

As a result of the calculations we obtained 8,9,10,11,12 and
13 pulse sequences. The sequences obtained and their characteristics
are summarized in Table \ref{tab:1}. Table  \ref{tab:1} also shows the characteristics
of the optimal 7-pulse sequence obtained in \cite{Farley_1972}, as
well as 7- and 8-pulse sequences used in the SuperDARN method\cite{Greenwald_et_al_1983,Riberio_et_al_2013}.
SuperDARN 13-pulse sequence \cite{Greenwald_et_al_2008} was not considered
in the comparison because it has a number of lags with a spatial resolution
worse than the elementary pulse duration.

Table \ref{tab:1} shows that the standard SuperDARN 7-pulse sequence is close to the optimum \cite{Farley_1972}, while
the katscan SuperDARN 8-pulse sequence is much less effective than
8 and 9 and the pulse sequence calculated by us. For a more detailed
comparison of these sequences at Figure 2 we show plots of
$d(T)$ for these sequences. From these graphs it is clear that the
sequences obtained by us have more 'good' lags than 7- and 8-pulse
sequences used previously.

Figure 2 shows that the found 9-pulse sequence (EKB 36-44) has almost
the same total duration as used now by 8-pulse katscan sequence \cite{Riberio_et_al_2013},
but the average power emitted by the radar in this case is 1.125 times
more, which means it also more efficient in terms of energy. From
the point of view of significant lags, the found 9-pulse sequence
has 28\% more significant 'good' lags than 8-pulse katscan, and its
first 'bad' lag is 36 instead of 6 for katscan. This makes 9-pulse
EKB 36-44 more efficient for measurements than the standard 8-pulse
katscan.

\begin{figure}
\includegraphics[scale=0.5]{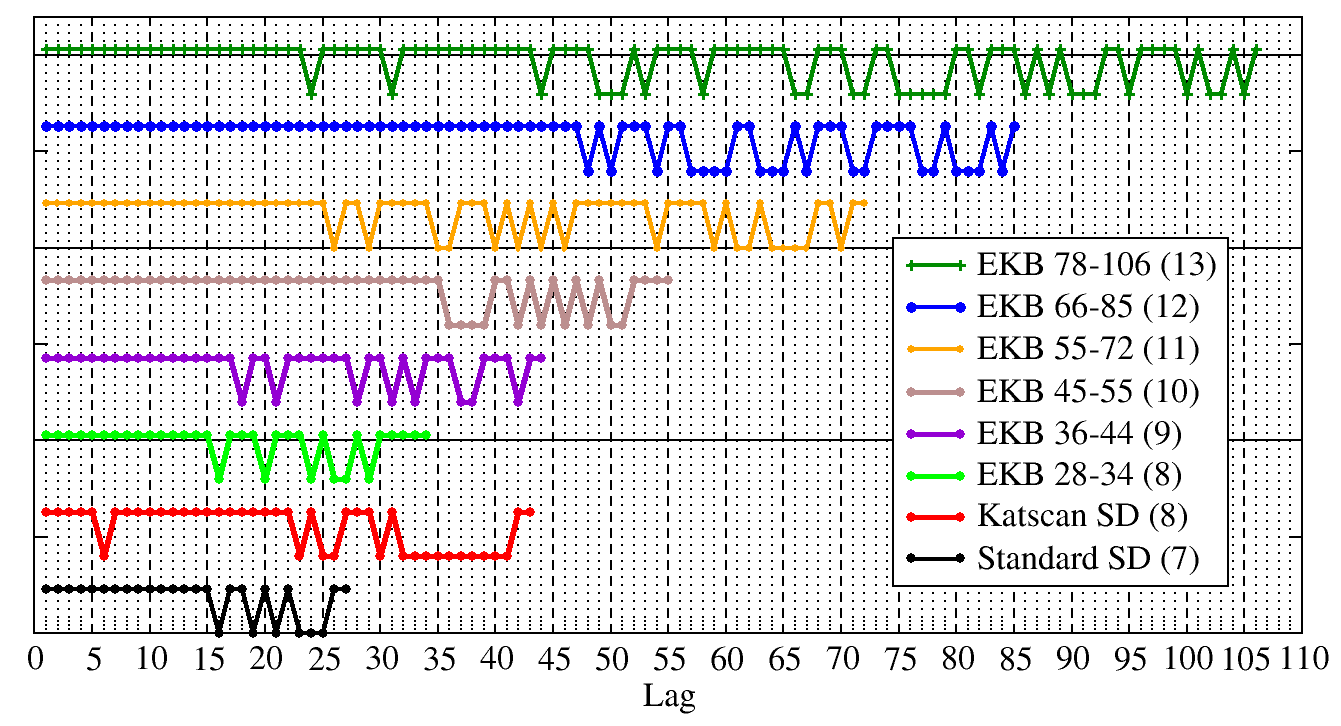}

\caption{Comparison of spatial resolution at different lags (i.e., the
position of good and bad lags) for different signals (from the bottom
up - a standard 7-pulse SuperDARN signal \cite{Greenwald_et_al_1983},
8-pulse katscan SuperDARN signal \cite{Riberio_et_al_2013}, 8,9,10,11
, 12 and 13 pulse signals found numerically)
}
\label{fig:2} 
\end{figure}

As can be seen from Table \ref{tab:1} and Fig.\ref{fig:2}, efficiency $\xi$ for the found 10-13-pulse sequences also close to the standard 7-pulse sequence
efficiency, but allow to analyze more lags in the correlation function,
and the first bad lag is always bigger than the first bad
lag of standard 7-pulse sequence.

Analysis of the shape of the weight volume shows that the requirement
of a minimum spatial resolution at all points is consistent with the
requirement of uniqueness interpulse intervals in the sounding sequence.
This requirement corresponds to the well-studied object in mathematics
- Golomb rulers \cite{Golomb_1972,Greenwald_et_al_2008}.

Analysis has shown that optimality criterion for SuperDARN sounding
signal we have formulated (the maximum number of good lags in the
weight volume) meets the criterion of optimality of Golomb rulers
(the maximum number of measurable numbers), and the task of finding
the optimal sequence thus reduces to the well-known and developed
problem of finding the optimal Golomb rulers.

For Golomb rulers to date optimal sequences up to
27 are numerically found, and the proof is kept now for the optimal power 28 sequences by the network computing project OGR-28 \cite{DistibutedNet_project}.
Optimal Golomb rulers are widely used in various practical problems. 

Table \ref{tab:2} shows the sounding signals corresponding to optimal
Golomb rulers with $N>6$, as well as their characteristics - number
of all lags L and good lags M in correlation function, the efficiency
$\xi$ and the average signal power P.

At table \ref{tab:2} we show the first bad lag for each of the sequences. The
table shows that the maximum length of part without bad lag that increases
monotonically with increasing number of pulses is provided by 8,9,10,12,14,18,20,24
and 26 pulse sequences. The rest of the sequences is much more worse.

Nearly optimal Golomb sequences differ from the optimal sequences
by slightly larger length and may also be used as a basis for sounding
signals.

We will refer the sounding sequences constructed from the condition of maximal
first bad lag in the class of optimal and nearly optimal \cite{Feiri} Golomb sequences
($dL <10\%$) as quasioptimal ones. We will refer the part of these
sequences that correspond to the optimal Golomb sequences as optimal
ones. Quasioptimal sequences provide a maximum of the first bad lag,
but do not always have the minimal possible length. Optimal sequence
realize the maximal first bad lag and minimal length at the same time.

Table \ref{tab:3} shows the quasioptimal sequences for $N<26$. An asterisk denotes
optimal sequences. For the values of 21-23 and 25,27 quasioptimal
sequences are not listed due to lack of available data. 

Table \ref{tab:3} shows that the standard 7-pulse SuperDARN signal is quasioptimal 7-pulse sequence in the framework of this approach, but 8-pulse katscan
SuperDARN signal, as expected from Table  \ref{tab:1}, is not quasioptimal signal. 

If not to pay much attention to the relative amount of good lags
in the correlation function, but only to the distance to the first
bad lag, any of the quasioptimal sequences (Table  \ref{tab:3}) can be used
for spectral measurements.

As one can see from (Table  \ref{tab:3}) the sequence providing maximal relative 
efficiency over all the parameters (maximal relative number of good lags 
and maximal relative position of first bad lag) is 10-pulse optimal Golomb ruler (Table  \ref{tab:3}).

Briefly, two problems of using new sounding signals must be taken into account. 

The first problem is reducing the signal/noise ratio which decreases
as the ratio of the number of pulses in a sequence to its full length
due to signals scattered from the uncorrelated ranges. 

The second problem is more significant for SuperDARN radars, is consists in
the statistical nature of the sounding, and so in order to determine the
parameters of scattered signal one need statistical averaging. The
accumulation time of the averaged signal for a fixed number of soundings
is proportional to the total sequence length, which actually leads
to increase in the required accumulation time as a square of the number
of pulses in the sequence.

Therefore, when choosing of the sounding sequence (increasing the
number of pulses in the sounding sequence) researcher is limited by required time resolution and radar energy. Since the scattering cross-section
of the irregularities investigated by SuperDARN radars is much higher
than the cross section of incoherent scattering, we should expect
that the optimal number of pulses in the sounding sequence may be greater
than 7 suggested in \cite{Farley_1972}, and we can use for the sounding
more complex signals obtained in this work. 

As can be seen from the above that from the optimal and nearly optimal
Golomb rulers the quasioptimal sounding SuperDARN signals can be chosen.
From these quasioptimal signals one can select the signal optimal
for researcher based on the necessary temporal resolution and the
energetic potential of the radar. For comparison, when using 12-pulse
quasioptimal sequence it is necessary to accumulate 3 times longer
than for the standard 7-pulse SuperDARN signal, by taking into account
the increased signal length. For a 10-pulse sequence accumulation
time increases only twice.

\section{Experimental results with quasioptimal sounding sequences}

On the basis of optimal and nearly optimal Golomb rulers, we have
constructed quasioptimal SuperDARN sounding signals for the number
of pulses from 7 to 26 (Table \ref{tab:3}), which simultaneously provide a relatively
high efficiency $\xi$ (differing from the optimal by not more than
10\%) and maximal first bad lag.

To verify the assumption of the effectiveness of obtained long sounding
sequences for SuperDARN radars and to obtain the first spectra of
scattered signals with high spectral resolution, we conducted experiments
at the EKB radar of the Russian segment of SuperDARN radars.

As showed the preliminary analysis, from the spectral measurements point
of view the optimal sounding sequences for SuperDARN (providing at
the same time the maximal relative amount of good lags and maximal
first bad lag) seems to be 10,12,14,18,20,24 and 26-element Golomb
rulers (Table  \ref{tab:3}) . However, with increasing of the pulses number in
the sequence significantly increases the required averaging time,
so experimental verification was carried out only for relatively
short 8, 9, 10 and 12-pulse sounding sequences. The first two of them
- are quasioptimal, the last two - optimal ones.

Standard formulation of the experiment was to switch every 12 minutes
between the 7,8,9,10 and 12 quasioptimal sounding sequences (Table
 \ref{tab:3}). In this case, the used 7-pulse sequence at the same time is a
standard SuperDARN signal.

Compared to the standard mode of EKB radar, the accumulation time
was increased twice (from 4 up to 8 seconds) to compensate for the
increased duration of the sounding signal. To calculate the spectra
we used the Wiener-Khinchin theorem, the discrete Fourier transform
with a window of 1000 samples and step interpolation of the correlation
function at bad lags. Further, all the spectral data will be given
in doppler velocity units (m/s), which is quite a standard representation
for SuperDARN radar data.

The experimental results are shown below.

To compare parameters obtained by spectral and correlation processing,
the spectra were processed by the following simple algorithm: to reduce
the influence of concentrated interference the
average spectral power averaged over all radar gates (distances) is
subtracted from spectral power at each range gate; as the signal power
the maximal spectral amplitude were used; as the average Doppler drift
the first moment of the spectrum was used; as the average spectral
width the second central moment of the spectrum was used; calculations
of spectral parameters were carried out in a range of speeds +/- 900m/s
to reduce the impact of concentrated interference at the 1000 m/s
(see Fig.\ref{fig:1}).

\begin{figure}
\includegraphics[scale=0.6]{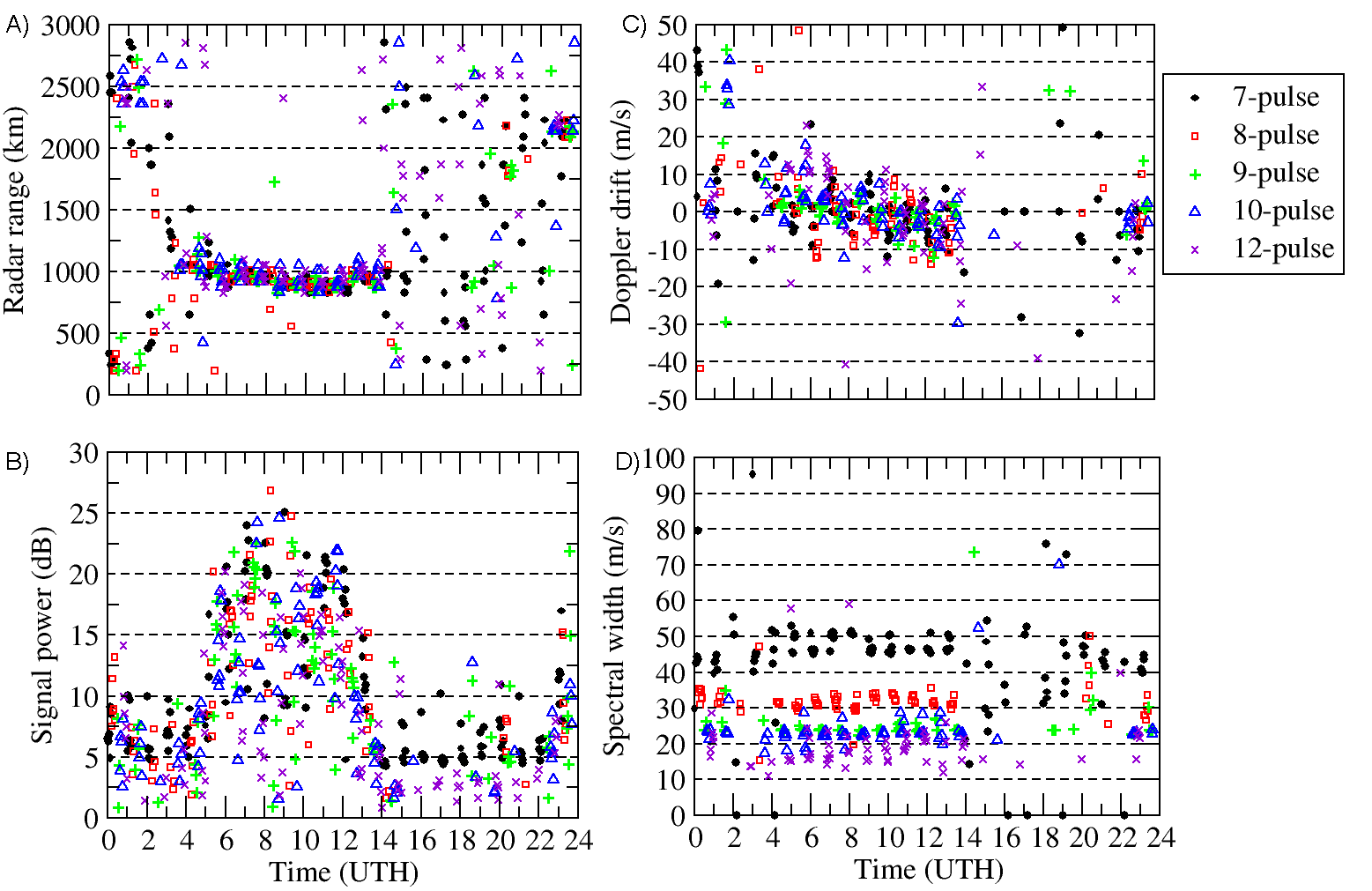}

\caption{
Characteristics of the signals at each time point (at north
direction), computed from the spectra of the signal. Range (A), power (B),
speed (C), and the spectral width (D) measured by 7,8,9,10 and 12-pulse
sequences. The signal is taken at ranges that provides the maximal
power of the scattered signal.
}
\label{fig:3} 
\end{figure}

Comparison of dynamics of these parameters is shown at Fig.\ref{fig:3}. The
comparison shows that the distributions of ranges with maximal signal
amplitude, the maximal power and Doppler drift for this range measured
in the experiments are quite similar, and weakly dependent on the
type of signal, which indicates validity of the results obtained from
the scattered signal spectrum. Spectral width is reduced with increasing
number of pulses in the sequence.

\begin{figure}
\includegraphics[scale=0.6]{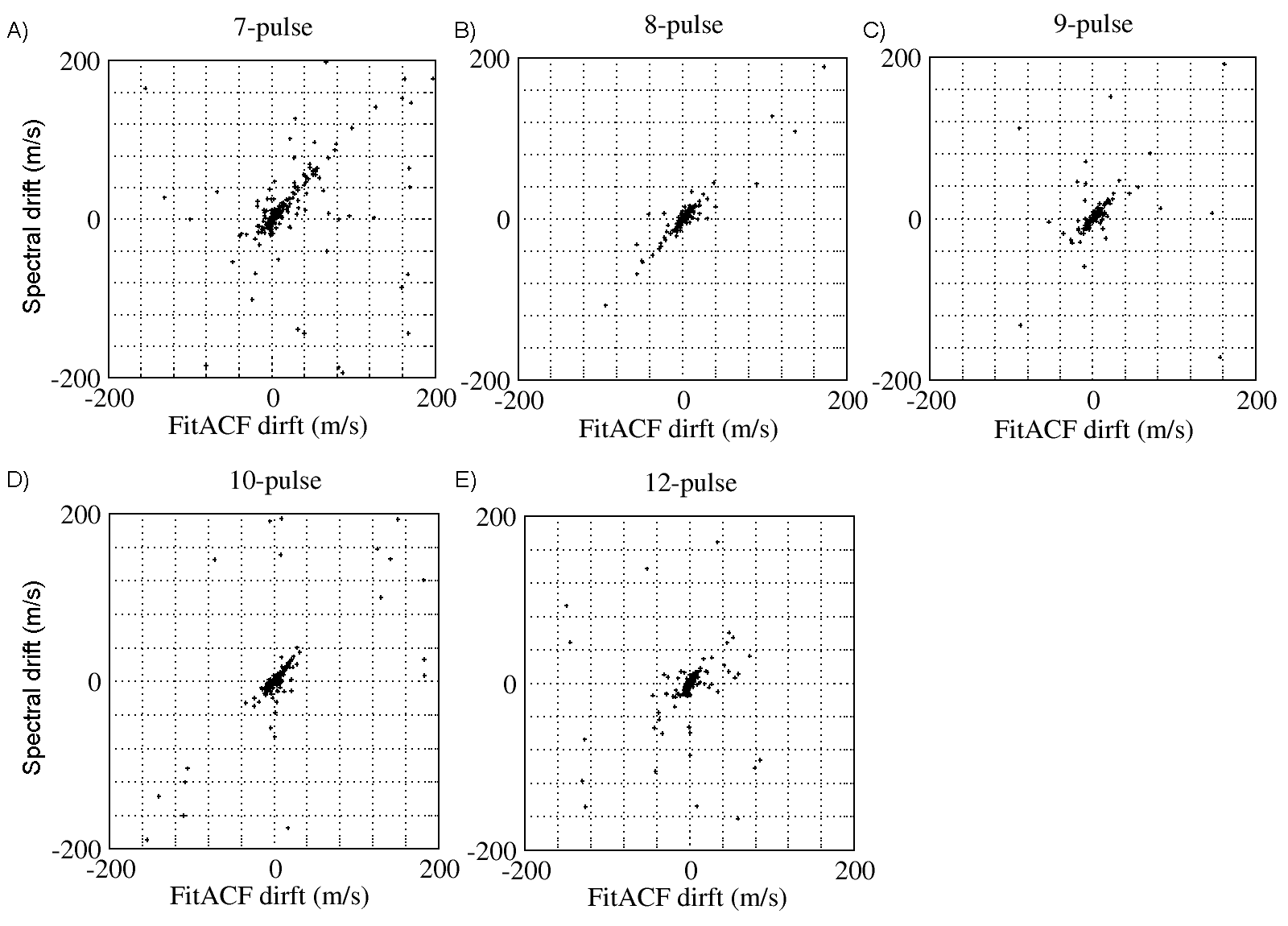}

\caption{
Comparison of Doppler frequency shift calculated by the standard
FitACF technique and by spectral method for a standard 7-pulse sequence (A)
and quasioptimal 8 (B),9 (C),10 (D) and 12-pulse (E) sequences.
}
\label{fig:4} 
\end{figure}

At Fig.\ref{fig:4} we show the velocities obtained by FitACF technique
and spectral method. When comparing the velocity the comparison was
made over the points, at which the FitACF technique prodices the correct
result with signal-to-noise ratio $>1$. One can see the average linear
proportionality of data obtained in the case of 7, 8,9,10 pulse sequences,
indicating the qualitative applicability of FitACF to interpret the
data obtained by new signals, and the correctness of the signals themselves.
Similar results were obtained when we use the LMFIT technique.

\begin{figure}
\includegraphics[scale=0.6]{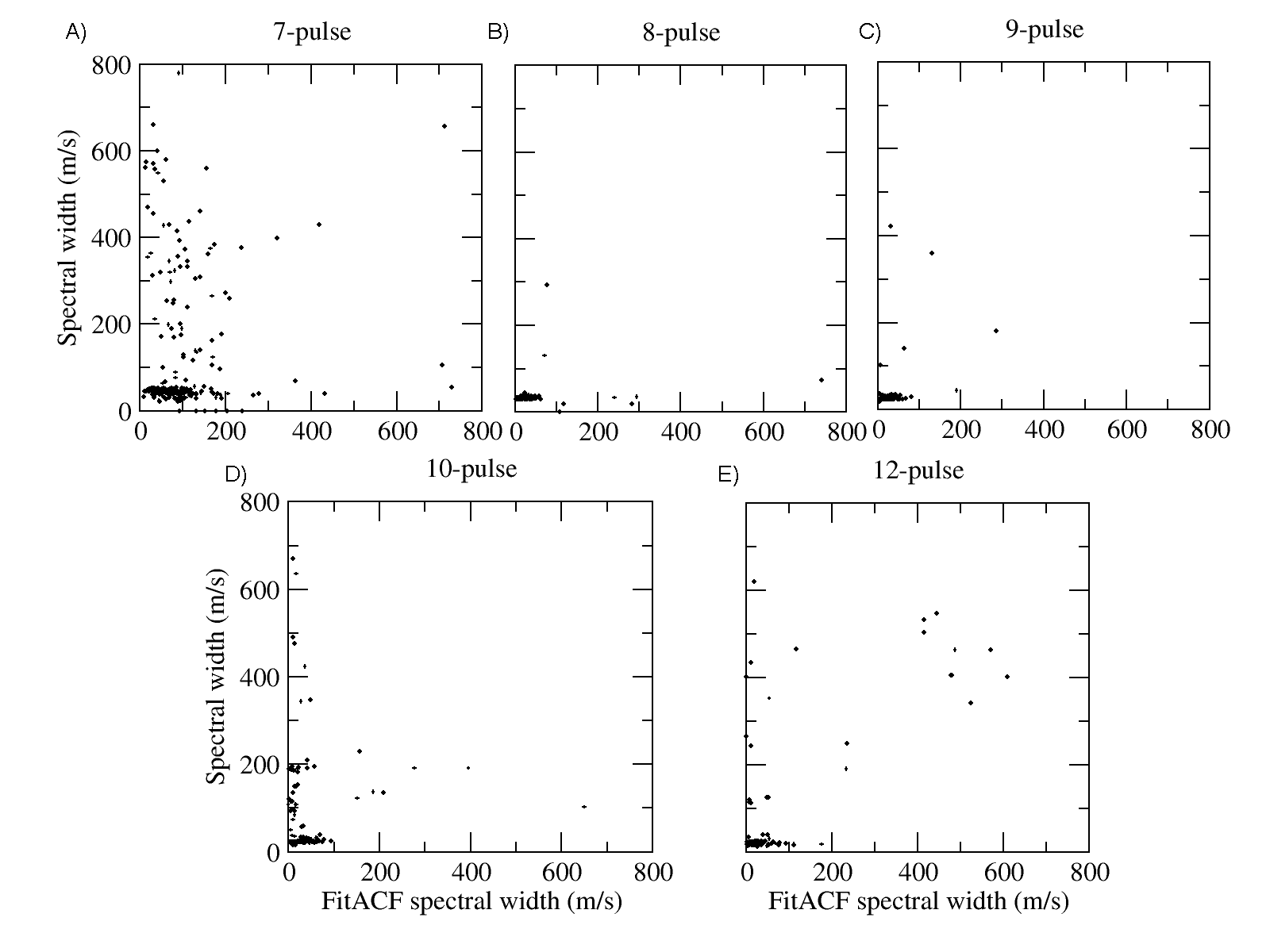}

\caption{
Comparison of the broadening of the Doppler spectrum calculated
by the standard FitACF technique and by spectral method for a standard
7-pulse sequence (A) and quasioptimal 8 (B),9 (C),10(D) and 12-pulse (E) sequences.
}
\label{fig:5} 
\end{figure}

At Fig.\ref{fig:5} we show a similar comparison between the spectral widths,
calculated by the standard FitACF technique and spectral methods.

At the fig.\ref{fig:5} one can see the area of greatest concentration of values
that are related to the constant spectral width of the spectrum at
a level correspondent to the theoretical spectral broadening for a
given sequence. At the same the spectral width of these signals, calculated
by the standard FitACF technique, differs significantly from the calculated
from the spectrum. It should be noted that in some cases, especially
for 7, 10 and 12-pulse sequences the high spectral widths calculated by
FitACF technique, are close to the spectral widths calculated from
the shape of the spectrum. This proves that high spectral widths are 
calculated well by FitACF technique for new pulses too.

\begin{figure}
\includegraphics[scale=0.4]{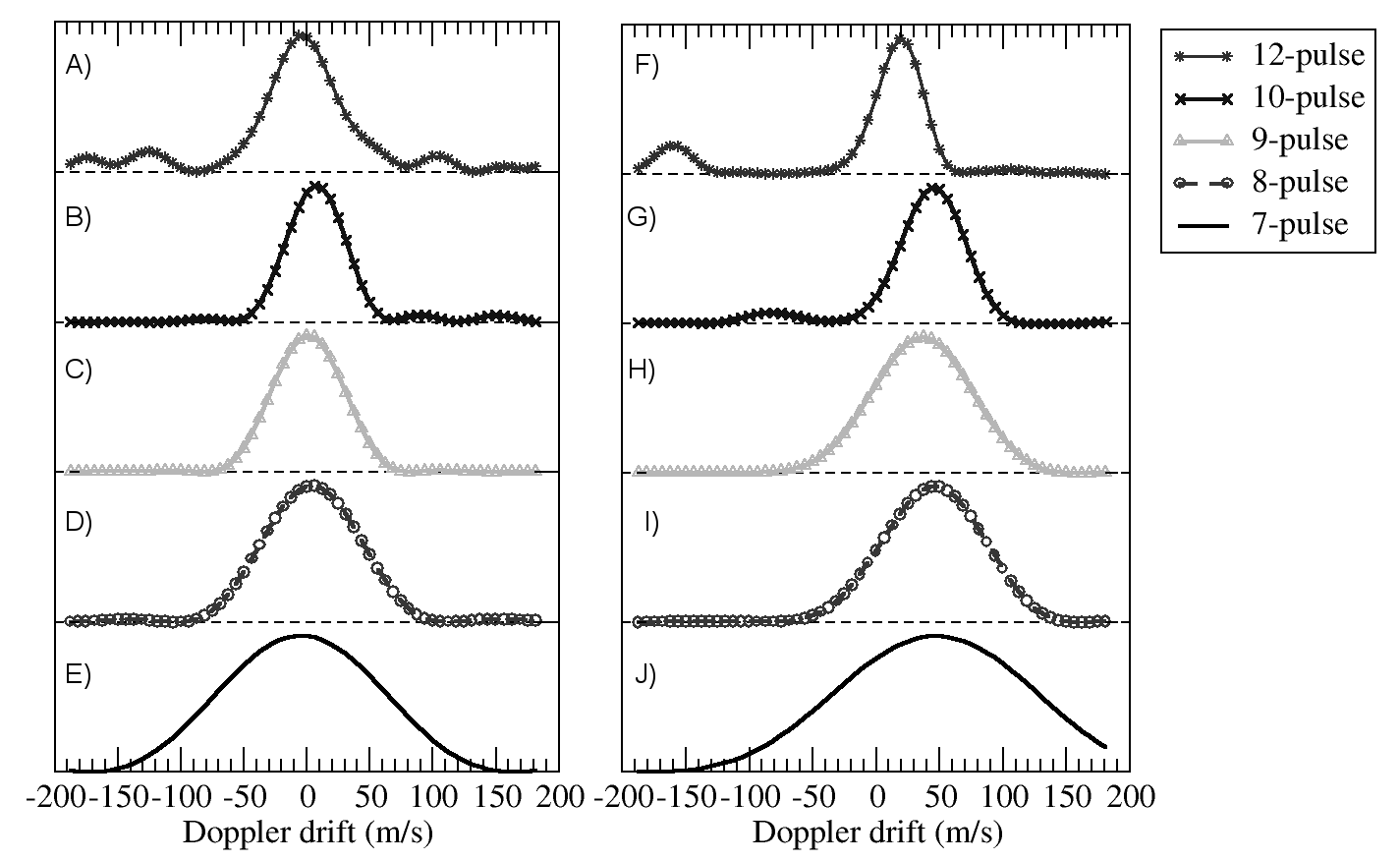}

\caption{Comparison of the scattered signal spectra in the case of ground
backscatter (A) at 09:00-10:00UT, R = 920km and scatter from ionospheric
irregularities (B) at 00:00-01:00UT, R = 1100km for the various shapes
of the sounding signal. Frequency offset is converted into an equivalent
Doppler velocity. Measurments with different signals are separated by 12 minutes.
}

\label{fig:6} 
\end{figure}

At Fig.\ref{fig:6} we show the comparison of the scattered signal spectra
in the case of the ground backscatter signal (Fig.\ref{fig:6}A) and ionospheric
scatter (Fig.\ref{fig:6}B) for the case of the standard 7-pulse sequence and
the new quasioptimal 8,9,10 and 12-pulse sequences. Ranges are identical,
measurement time is approximately the same (with 1 hour accuracy).
Accumulation time 8 seconds. From Fig.\ref{fig:6}A one can see that there is
an effective narrowing of the scattered signal spectrum, as it is
expected for using of a longer sounding sequence and is characteristically
for the ground backscatter signal, that is narrowband enough. Doppler
shift of the signal at the same time is less than 20m/s, which is
also quite typical for ground backscatter signal. Sometimes, for example
in the case of a 12-pulse sequence (Fig.\ref{fig:6}A) the spectrum has asymmetrical
form, indicating the possible presence of multiple modes in the received
signal. Thus, the ground backscatter signal showed that the parameters
obtained by new sounding signals are in qualitative agreement (low
spectral width, small Doppler drifts) with the expected characteristics.

From Fig.\ref{fig:6}B one can see that sometimes the signal scattered by ionospheric
irregularities has an asymmetric spectra, sometimes with weak additional
peaks (as well seen for 12- and 10-pulse sequences at Fig.\ref{fig:6}B)). Standard
7-pulse sequence in spectral technique is less sensitive to these
effects \cite{Danskin_et_al_2004}, and therefore it is necessary to use more sophisticated inversion
techniques \cite{Barthes_et_al_1998,Danskin_et_al_2004,Riberio_et_al_2013} for the analysis
of cases with complex or asymmetric spectra of the scattered signal.

\section{Conclusion }

In this paper we formulate a criterion of quasioptimal SuperDARN sounding
signals for spectral measurements, which consists from simultaneous
requirements: to maximize the first bad lag number in the correlation
functions of the received signal; to decrease the density of bad lags,
close to the absolute minimum; to provide spatial resolution that not worser
than elementary pulse duration at all delays except zero. Solution
of the problem corresponds to finding Golomb rulers \cite{Golomb_1972},
satisfying the specified conditions.

On the basis of optimal and nearly optimal Golomb rulers we constructed
quasioptimal sounding sequences satisfying the formulated principles,
for the number of elementary pulses 7-20,24,26 (Table  \ref{tab:3}). It is shown
that according to this approach the standard SuperDARN 7-pulse sounding signal \cite{Greenwald_et_al_1983}
is quasioptimal one, and 8-pulse katscan SuperDARN
sounding signal \cite{Riberio_et_al_2013} is not quasioptimal one.

It is shown that the optimal (for which all the required conditions
are met, and the length of sequence becomes the minimal possible)
from this point of view are 10, 12, 14, 18, 20  and possibly
24 and 26-pulse optimal Golomb rulers (Table  \ref{tab:3}). 

It is shown that the sequence providing maximal relative efficiency over all the parameters (maximal relative number of good lags 
and maximal relative position of first bad lag) is 10-pulse optimal Golomb ruler (Table  \ref{tab:3}).

Experiments were made for sounding by the quasioptimal 8, 9, 10 and
12-pulse sequences at EKB radar and it is shown the continuity of
the obtained data with data obtained by the standard 7-pulse sequence.
We also obtained the expected improvement in spectral resolution.

\section*{Acknowledgments}

The authors are grateful to ISDCT SB RAS for providing computing clusters
ISC SB RAS - 'Blackford' and 'Academician
V.M.Matrosov' (http://hpc.icc.ru) for numerical calculations.
The authors thank Michael Feiri (University of Twente, Enschede, Netherlands)
for the results of the calculation of nearly optimal Golomb sequences
for $N<21$ (http://www.feiri.de/ogr/nearopt.html) and 
Pavlo Ponomarenko (University of Saskatchevan, Saskatoon, Canada) for useful discussions.

\newpage

\begin{table}
\begin{tabular}{|p{5cm}|c|p{5cm}|c|c|}
\hline 
Type & N & Pulse positions & $\xi$ & First bad lag \\
\hline 
\hline 
{\scriptsize{21-25 (\cite{Farley_1972}) }} & {7} & {0 1 11 16 19 23 25} & {0.84} & {13} \\
\hline 
{\scriptsize{21-27 (Standard SD, \cite{Chisham_et_al_2007}) }} & {7} & {0 9 12 20 22 26 27 } & {0.77} & {16} \\
\hline 
{\scriptsize{28-43 (Katscan SD, \cite{Riberio_et_al_2013}) }} & {8} & {0 1 12 16 19 21 29 43 } & {0.64} & {6} \\
\hline 
{\scriptsize{EKB 28-34 (this paper)}} & {8} & {0 1 4 9 15 22 32 34 } & {0.82} & {16} \\
\hline 
{\scriptsize{EKB 36-44 (this paper)}} & {9} & {0 1 5 12 25 27 35 41 44} & {0.82} & {18} \\
\hline 
{\scriptsize{EKB 45-55 (this paper)}} & {10} & {0 1 6 10 23 26 34 41 53 55 } & {0.82} & {36} \\
\hline 
{\scriptsize{EKB 55-72 (this paper)}} & {11} & {0 1 9 19 24 31 52 56 58 69 72} & {0.76} & {26} \\
\hline 
{\scriptsize{EKB 66-85 (this paper)}} & {12} & {0 2 6 24 29 40 43 55 68 75 76 85} & {0.76} & {48} \\
\hline 
{\scriptsize{EKB 78-106 (this paper)}} & {13} & {0 2 5 25 37 43 59 70 85 89 98 99 106} & {0.73} & {24} \\
\hline 
\end{tabular}

\caption{Comparison of the standard sounding sequences used at SuperDARN
radars and numerically found more powerful sequences}
\label{tab:1}

\end{table}

\newpage

\begin{table}

~

\begin{tabular}{|c|c|c|p{7cm}|c|c|c|}
\hline 
N & M & L & Pulse positions & $\xi$ & First bad lag & P \\
\hline 
\hline 
{7} & {21} & {25} & {0 2 3 10 16 21 25} & {0.84} & {13} & {0.28} \\
\hline 
{8} & { 28} & {34} & {0 1 4 9 15 22 32 34} & {0.82} & {16} & { 0.23} \\
\hline 
{9} & { 36} & {44} & {0 1 5 12 25 27 35 41 44} & {0.82 } & {18} & {0.2} \\
\hline 
{10+*} & { 45} & { 55} & {0 1 6 10 23 26 34 41 53 55} & {0.82} & {36} & { 0.18} \\
\hline 
{11} & { 55} & { 72} & {0 1 9 19 24 31 52 56 58 69 72} & {0.76} & {26} & {0.15} \\
\hline 
{12+* } & { 66} & {85} & {0 2 6 24 29 40 43 55 68 75 76 85} & {0.78} & {48} & { 0.14} \\
\hline 
{13 } & { 78} & {106} & {0 2 5 25 37 43 59 70 85 89 98 99 106} & {0.73} & {24} & { 0.12} \\
\hline 
{14* } & { 91} & {127} & {0 4 6 20 35 52 59 77 78 86 89 99 122 127} & {0.72} & {56} & { 0.11} \\
\hline 
{15 } & { 105} & {151} & {0 4 20 30 57 59 62 76 100 111 123 136 144 145 151} & {0.69} & {18} & { 0.10} \\
\hline 
{16*} & { 120 } & {177} & {0 1 4 11 26 32 56 68 76 115 117 134 150 163 168 177 } & {0.68} & {23} & { 0.09} \\
\hline 
{17 } & { 136 } & {199} & {0 5 7 17 52 56 67 80 81 100 122 138 159 165 168 191 199} & {0.68} & {18} & {0.08} \\
\hline 
{18+*} & { 153 } & {216 } & {0 2 10 22 53 56 82 83 89 98 130 148 153 167 188 192 205
216} & {0.70} & {91} & {0.08} \\
\hline 
{19 } & {171} & {246} & {0 1 6 25 32 72 100 108 120 130 153 169 187 190 204 231
233 242 246 } & {0.69} & {50} & {0.08} \\
\hline 
{20* } & {190} & {283 } & {0 1 8 11 68 77 94 116 121 156 158 179 194 208 212 228
240 253 259 283 } & {0.67 } & {98} & {0.07} \\
\hline 
{21 } & { 210} & {333} & {0 2 24 56 77 82 83 95 129 144 179 186 195 255 265 285
293 296 310 329 333} & {0.63} & {29} & {0.06} \\
\hline 
{22 } & { 231} & {356 } & {0 1 9 14 43 70 106 122 124 128 159 179 204 223 253 263
270 291 330 341 353 356} & {0.65 } & {24} & {0.06} \\
\hline 
{23+} & { 253} & {372 } & {0 3 7 17 61 66 91 99 114 159 171 199 200 226 235 246
277 316 329 348 350 366 372} & {0.68} & {62} & { 0.06} \\
\hline 
{24* } & { 276} & {425 } & {0 9 33 37 38 97 122 129 140 142 152 191 205 208 252 278
286 326 332 353 368 384 403 425} & {0.65} & {128} & { 0.06} \\
\hline 
{25 } & { 300} & {480 } & {0 12 29 39 72 91 146 157 160 161 166 191 207 214 258
290 316 354 372 394 396 431 459 467 480} & {0.62} & {81} & {0.05} \\
\hline 
{26+*} & { 325} & {492 } & {0 1 33 83 104 110 124 163 185 200 203 249 251 258 314
318 343 356 386 430 440 456 464 475 487 492} & {0.66} & {159} & {0.05} \\
\hline 
{27 } & { 351} & {553 } & {0 3 15 41 66 95 97 106 142 152 220 221 225 242 295 330
338 354 382 388 402 415 486 504 523 546 553} & {0.63} & {32} & {0.05} \\
\hline 
\end{tabular}

\caption{Characteristics of sounding signals based on optimal Golomb
rulers. Signals that provide local optimal maximum of $\xi$ are marked
by crosses, signals that provide local maximum of the first bad lag
are marked by asterisks.
}
\label{tab:2} 

\end{table}

\newpage

\begin{table}

\begin{tabular}{|c|c|c|p{7cm}|c|c|c|}
\hline 
{N} & {M} & {L} & {Pulse positions} & {$\xi$} & {First bad lag} & {FBL/L} \\
\hline 
\hline 
{7} & {21} & {27} & {0 9 12 20 22 26 27 } & {0.77} & {16} & {0.59} \\
\hline 
{8} & {28} & {35} & {0 4 5 17 19 25 28 35} & {0.80} & {22} & { 0.62} \\
\hline 
{9} & {36} & {45} & {0 2 10 24 25 29 36 42 45} & {0.80} & {28} & {0.62} \\
\hline 
{10*} & {45} & {55} & {0 1 6 10 23 26 34 41 53 55} & {0.82} & {36*} & { 0.65} \\
\hline 
{11} & {55} & {83} & {0 4 5 16 22 24 31 45 70 73 83} & {0.66} & {30} & {0.36} \\
\hline 
{12*} & {66} & {85} & {0 2 6 24 29 40 43 55 68 75 76 85} & {0.78} & {48*} & {0.56} \\
\hline 
{13} & {78} & {113} & {0 3 13 25 33 39 54 85 86 102 104 109 113} & {0.69} & {34} & { 0.30} \\
\hline 
{14*} & {91} & {127} & {0 4 6 20 35 52 59 77 78 86 89 99 122 127} & {0.72} & {56*} & { 0.44} \\
\hline 
{15} & {105} & {156} & {0 2 15 33 49 56 68 104 118 126 129 146 150 155 156} & {0.67} & {39} & { 0.25} \\
\hline 
{16} & {120} & {184} & {0 11 32 33 60 68 75 77 80 106 131 147 161 165 171 184 } & {0.65} & {39} & { 0.21} \\
\hline 
{17} & {136} & {201} & {0 5 15 34 35 42 73 75 86 89 98 134 151 155 177 183 201} & {0.67} & {72} & {0.35} \\
\hline 
{18* } & {153} & {216} & {0 2 10 22 53 56 82 83 89 98 130 148 153 167 188 192 205 216} & {0.70} & {91*} & {0.42} \\
\hline 
{19} & {171} & {294} & {0 3 7 13 28 39 47 48 98 121 167 184 198 200 222 227 264 276 294} & {0.58} & {52} & {0.17} \\
\hline 
{20* } & {190} & {283} & {0 1 8 11 68 77 94 116 121 156 158 179 194 208 212 228 240 253 259 283 } & {0.67} & {98*} & {0.34} \\
\hline 
{24* } & {276} & {425} & {0 9 33 37 38 97 122 129 140 142 152 191 205 208 252 278 286 326 332 353 368 384 403 425} & {0.65} & {128*} & { 0.30} \\
\hline 
{26*} & {325} & {492} & {0 1 33 83 104 110 124 163 185 200 203 249 251 258 314 318 343 356 386 430 440 456 464 475 487 492} & {0.66} & {159*} & {0.32} \\
\hline 
\end{tabular}

\caption{Characteristics of quasioptimal sounding signals on the basis
of optimal and nearly optimal Golomb rulers ($dL <10\%$). An asterisk
denotes the optimal sequences.
The last column - relation of first bad lag position to total number of lags
}
\label{tab:3} 

\end{table}

\newpage
\newpage

\newpage

\newpage

\newpage

\newpage

\newpage

\end{document}